# A versatile and reproducible cryo-sample preparation methodology for atom probe studies

Eric V. Woods[1,#,*], Mahander P. Singh[1,#], Se-Ho Kim[1], Tim M. Schwarz[1], James O. Douglas[2], Ayman El-Zoka[1,2], Finn Giulani[2], and Baptiste Gault[1,2,*]

1. Max-Planck-Institut für Eisenforschung GmbH, Max-Planck-Straße 1, 40237 Düsseldorf, Germany

2. Department of Materials, Royal School of Mines, Imperial College London, Prince Consort Road, London SW7 2BP, UK.

# Authors have contributed equally to this article

*Corresponding Authors: ewoods@mpie.de, b.gault@mpie.de

## Abstract

Repeatable and reliable site-specific preparation of specimens for atom probe tomography (APT) at cryogenic temperatures has proven challenging. A generalized workflow is required for cryogenic-specimen preparation including lift-out via focused-ion beam and in-situ deposition of capping layers, to strengthen specimens that will be exposed to high electric field and stresses during field evaporation in APT, and protect them from environment during transfer into the atom probe. Here, we build on existing protocols, and showcase preparation and analysis of a variety of metals, oxides and supported frozen liquids and battery materials. We demonstrate reliable in-situ deposition of a metallic capping layer that significantly improve the atom probe data quality for challenging material systems, particularly battery cathode materials which are subjected to delithiation during the atom probe analysis itself. Our workflow designed is versatile and transferable widely to other instruments.

## 1. Introduction

Atom probe tomography (APT) offers a unique combination of high spatial resolution and elemental sensitivity in the atomic part-per-million range, to provide, three-dimensional analysis of the distribution of atoms in a material (Gault et al., 2021; Lim et al., 2020; De Geuser & Gault, 2020; Devaraj et al., 2018). The overall process involves generating high electric fields at the end of a needle-shaped specimen, typically less than 100 nm in diameter, that cause the joint ionization and desorption of atoms from the surface, triggered by either laser or voltage pulses (Gault et al., 2006; Hono et al., 2011). These needle shaped specimens, are routinely prepared using dual-beam scanning-electron microscopes-focused-ion beams (SEM-FIB) (Thompson et al., 2007; Prosa & Larson, 2017). However, the FIB-based preparation of beam-sensitive materials or liquid samples remain challenging (Schreiber et al., 2018; Bassim et al., 2012). Cryogenic freezing can preserve the sample's structure and composition, enabling high-quality analysis while minimizing FIB milling-induced damage (McCarroll et al., 2020; El-Zoka et al., 2020; Rivas et al., 2020). Development of cryogenic stages for the SEM-FIBs has enabled the analysis of wider range of materials using APT, and cryogenic shuttle systems allow for transfer of specimens between instruments, enabling APT analysis of materials that are sensitive to air or change in temperatures (Gerstl & Wepf, 2015; Perea et al., 2017; Stephenson et al., 2018).

These developments have facilitated APT studies of hydrogen in steel (Chen et al., 2017) and aluminum (Zhao et al., 2022), elemental distributions in battery materials (Singh et al., 2023; Li et al., 2022). And frozen liquids and their interface with a metallic substrate (Schreiber et al., 2018; Schwarz et al., 2020; El-Zoka et al., 2020; Kim, Stephenson, et al., 2022). Cryogenic techniques for preparing frozen liquid specimen for APT analysis are still under development, but they offer a promising approach for characterizing the distribution of species in solutions. However, further research and development is required to fully realize the potential of this approach.

One of the inherent difficulty of preparing specimens at cryogenic temperatures is the inability to use the gas injection system (GIS) used to weld and lift-out a region-of-interest (ROI), and mounting it on the support (Prosa & Larson, 2017). Conventional GIS are often unsuitable for cryogenic environments, as the organic and metalorganic vapors they produce tend to condense over the entire cooled surface (Parmenter et al., 2014; Perea et al., 2017), precluding facile localized welding.

In recent years, alternative approaches for preparing samples at cryogenic temperatures have been investigated. FIB-free methods have been developed, such as embedding the liquids between a metallic specimen and graphene layers (Qiu et al., 2020; Zhang et al., 2022), making specimens from frozen liquids on a flat substrate (El-Zoka et al., 2020) or on wire (Schwarz et al., 2020; Stender et al., 2022), or infiltrating the liquid into a porous structure (Kim, El-Zoka, et al., 2022). These techniques have shown promise but severe limitations with respect to the volume that can be encapsulated, the lack of control of the freezing rate and the impossibility to select a ROI.

FIB preparations of beam sensitive samples, including battery materials or Li containing alloys pose another set of challenges, including Ga-implantation, Ga damage, and associated damage along with the reactivity of Ga that can cause critical issues (Belkacemi et al., 2023; Santhanagopalan et al., 2014). Cryogenic preparations of APT specimens can be of great help in overcoming the issues and being air sensitive, use of transfer cryo-shuttles can facilitate the APT analysis (Kim, Dong, et al., 2022; Chang et al., 2019).

Along with the specimen preparation, APT analysis of frozen liquids faces issues often reported arising from localized heating at the field evaporating surface (McCarroll et al., 2020; Kelly et al., 2009). Recent reports suggest that the use of thin conductive coatings led to an increase in yield by using e.g. metals films or graphene (Larson et al., 2013; Seol et al., 2016; Adineh et al., 2018) suggesting that a free flow of charge at the surface of the specimen facilitates controlled field evaporation along with the heat extraction. Recent work also showed that a conductive coating on oxide materials was greatly improving APT data quality and even enabling some analysis (Kim, Antonov, et al., 2022; Kim et al., 2023; Seol et al., 2016). These approaches require exposition of the specimen to ambient conditions to insert them into a separate device for deposition of the coating. An in-situ coating was introduced by (Kölling & Vandervorst, 2009) by using redeposition inside the FIB, which was revisited by Douglas et al. who used it to reinforce welds between sample and Si support by sputtering the W micromanipulator at cryogenic temperature (Douglas et al., 2022) .

Here, building on these methods, we introduce a workflow for APT specimen preparation under cryogenic conditions, from a glovebox to a FIB for site-specific lift out, specimen sharpening and in-situ coating to an atom probe for analysis with cryogenic, UHV transfer. The workflow is demonstrated in the analysis of a ceramic, namely (anodized alumina membrane), a Cu-based alloy (de-zincified Brass), supported and unsupported frozen aqueous solutions, and finally a commercial battery cathode oxides (NMC811) following optimization of an in-situ capping protocol. Our approach can be used a wide range of microscopes, at cryogenic and room temperature, allows to protect specimens from the exposure to ambient environment, which makes it of general interest across the community.

## 2. Materials and Methods

### 2.1 Materials

For the present study, specimens were produced from a range of materials systems, the detail of the preparation are provided below in their respective sections.

Anodized alumina oxide (AAO) membrane with 80 nm pore size and 50 µm thickness was sourced from InRedox, Longmont CO, USA, electrodeposited with nickel following the procedure described in (Lim et al., 2020).

Yellow brass (Cu: 63 % & Zn: 37 %) sheets with 0.4 mm thickness, (Metall Ehrnsberger GbR, Teublitz Germany) were cut into 3 mmx 7 mm pieces, annealed for 1 hr and cooled under Argon. Frozen water on de-zincified brass substrate.

For liquids, a solution of 200 mL Type 1 deionized (DI) water (Sigma Aldrich, St. Louis, MO, USA) and 5 g L-Arginine hydrochloride (HCl) (Alfa Aesar, Kandel, Germany) was prepared. Phosphate buffered saline (PBS) solution, 0.1 mM concentration, was purchased from Sigma-Aldrich directly.

NMC811 ($LiNi_8Co_1Mn_1O_2$) battery cathode material, sourced from Targray (Kirkland, Canada).

### 2.2 Focused Ion Beam (FIB) with cryo-stage

A Thermo-Fisher Helios 5 CX Gallium FIB/SEM (Thermo-Fisher Scientific, Waltham MA, USA) equipped with an Aquilos cryo-stage with free rotation capability and a Thermo-Fisher EZ-Lift tungsten cryogenic manipulator was used for specimen preparation. The stage can be cooled to a set point of -190 $^o$C and the manipulator to -175 $^o$C by a circulation of $N_2$ gas flow of 190 mg/s through a heat exchanger system within a liquid nitrogen Dewar. To cool the stage and the micromanipulator to the target temperature, one hour is required to reach thermal equilibrium and avoid any drift during the subsequent preparation. In the following, all SEM images are shown with a red border, whereas ion-beam images are delineated in blue.

A Sylatech glovebox (Sylatech GmbH, Walzbachtal, Germany) was used to maintain a level of moisture normally less than 20 ppm oxygen, and a dew point of -98 °C, for handling air sensitive specimens and for plunge freezing into LN2. A dual post cryogenic atom probe puck (Cameca Instruments, Madison, WI, USA), e.g. one with two sitting holes, i.e. slots that accommodate each a specimen holder with Cu springs ("Cu clip", Cameca) in opposite directions. A commercial 36-microtip array Si coupon (Cameca FT 36) was used a support, and is mounted on a Cu clip in one of the slots. Images of the dual puck is provided in the Supplementary files Fig. S1.

## 2.3 Atom probe Tomography (APT)

The APT measurements were conducted using a local electrode atom probe (LEAP, Cameca Instruments Inc.), either straight flight path (LEAP 5000 XS) or reflectron (LEAP 5000 XR). Measurements were done at different conditions as per the samples. The reconstruction of the three-dimensional atom maps and data analysis was carried out using the commercial software AP Suite 6.1.

## 3. Cryo-lift out workflow

As a preliminary step of this workflow, a cuboidal bar having dimensions 20 x 2 x 12 µm was extracted from bulk Cr piece using a FIB protocol described in the Ref. (Giannuzzi & Stevie, 1999). Cr was selected because it is widely available, is often used as a coating material for SEM or APT specimens and was shown in previous work to exhibit good adhesion to flat and curved surfaces (Kölling & Vandervorst, 2009). The Cr-lamella was attached with the micromanipulator using Pt via. GIS system, and all steps were performed at room temperature, shown in Fig 1(a). This attached Cr-lamella will later be used during the subsequent stages of specimen preparation.

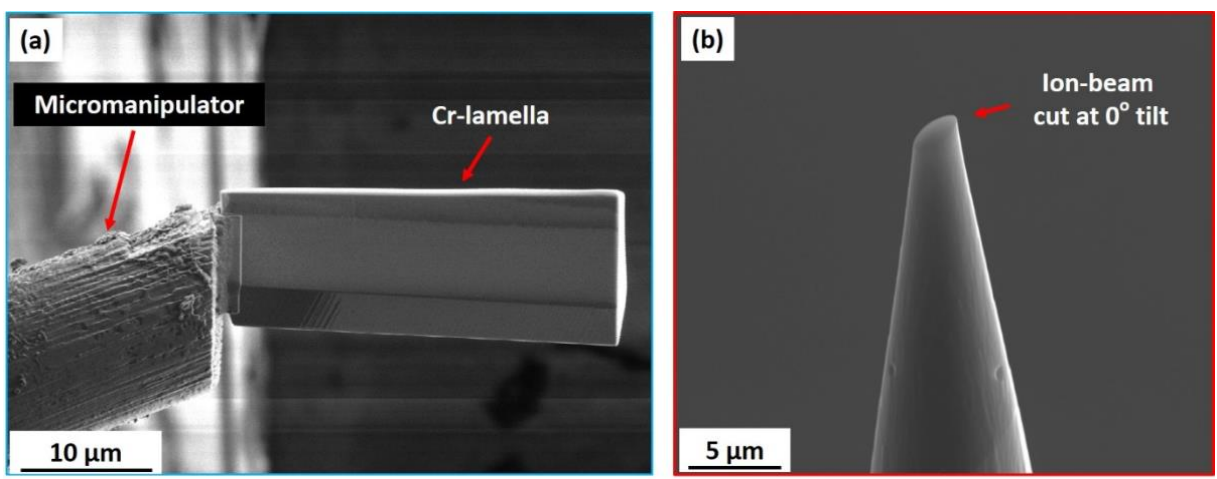

**Figure 1.** (a) Ion-beam image of the Cr-lamella lift-out, Pt welding was used and the lift was done at room temperature, (b) Secondary electron beam image of the sliced Si-micro post, milling was performed at 0° degree tilt.

This attachment could also be performed at cryogenic temperature using the protocol described by (Schreiber et al., 2018) if necessary. Another preliminary step, and to increase the contact area of Si posts and trenched lamella, the top part of the flat post of the 3 µm-wide flat top Si microtip arrays are sliced at 0° degree tilt, shown in Fig. 1(b).

For the cryo-lift-out, the stage was cooled to -190 °C, and the micro-manipulator was cooled to -175 °C. A schematic of the step-by-step cryo-lift out procedure is presented in Fig. 2. The first step of the process involves locating the ROI, and milling the adjoining regions leaving a ROI as a slice of 3 – 5 µm in thickness, very similar to those used for the well-established "lift-out" technique for TEM cross-section specimen preparation (Giannuzzi & Stevie, 1999), shown in Fig. 2(a). The sample is brought to 0° degree tilt to mill undercuts leaving the slice only attached to one edge. The micromanipulator attached with the Cr-lamella is then inserted and maneuvered close (<0.2 µm) to the trenched bar from ROI. At this point, cuts for redeposition (Schreiber et al., 2018) are used to attach the Cr-lamella to the ROI. 6 – 8 line cuts with 4 µm length and cut depts of 1 µm were made using ion-beam current of 80 pA at 30 kV, shown as white dashed lines in Fig. 2(b). The bar is then extracted from the bulk sample, leaving the ROI bar attached with Cr-lamella, as shown in Fig. 2(b).

In the next step, the bar is aligned to the top of pre-sliced Si posts. The bar is then secured to the post using vertical line cuts (4 – 5 lines) for redeposition, using ion-beam current of 80 pA at 30 kV, as shown with vertical white dashed lines in Fig. 2(c). Once the weld is secured, the ion beam is used to slice the remaining bar from the mounted slice. This process is repeated to prepare multiple slices from a single lifted-out bar, Fig. 2(c). Similar redeposition cuts are made from all sides by rotating the stage by 90°, at 0° tilt. This step fortifies the bond between the slice and the post.

In the next step, a rectangular cut is made at the interface, roughly removing half of the thickness of the mounted slice, Fig. 2(c). The Cr-lamella attached to the micromanipulator is maneuvered close to this cavity. Cr is ablated from the Cr-lamella using current of 80 pA at 30 kV using regular cross-section with cut depth of 1 µm, shown in yellow color in Fig. 2(d). Cr deposits in the cavity, at the interface between the sample and the Si post, provides a strong bond analogous to the Pt in the conventional FIB-based preparation for APT specimens.

Suitable needle shaped specimens are then obtained by applying a sequence of annular mill patterns to the specimen post in the FIB, where outer diameter of the mill pattern is slightly larger than the outer diameter of the sample and the inner diameter gets continuously smaller. The stage is tilted to 52° and annular milling is performed and the final tips in shape of sharp needles are obtained as shown in Fig. 2e.

Supplementary table T1 lists all the currents used for the lift out, which varies with the nature of the sample.

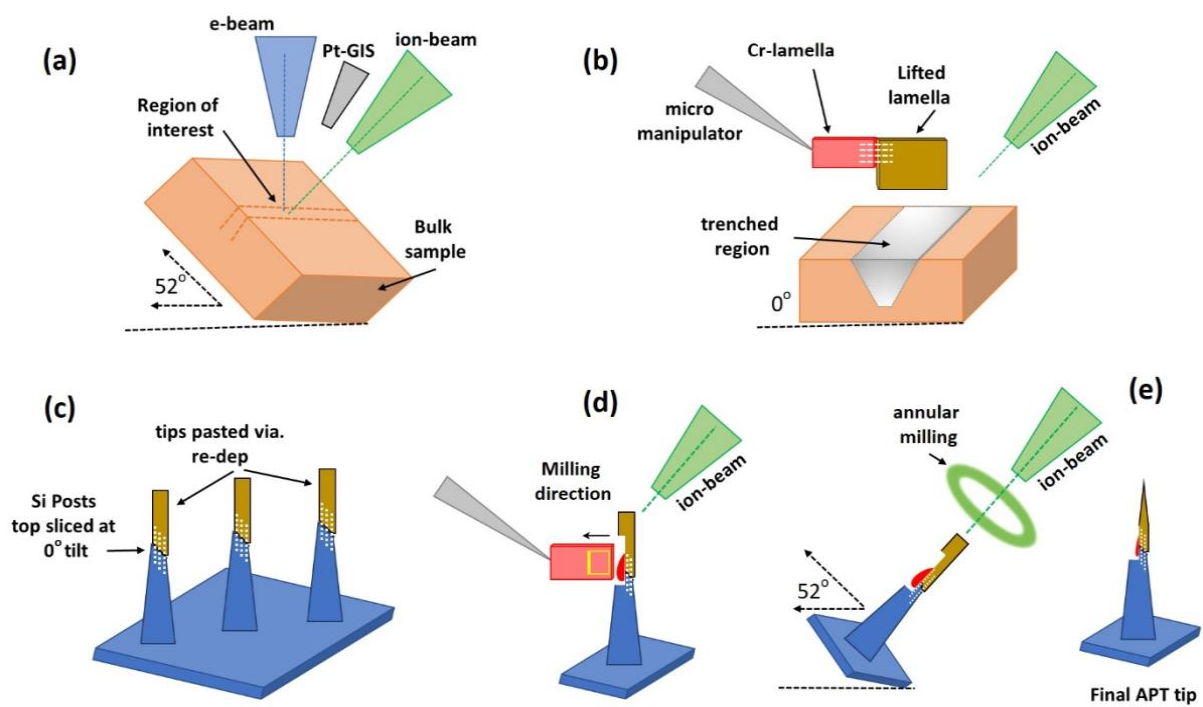

**Figure 2.** Schematic showing step-by-step preparation of cryo-FIB sample preparations, developed in this work.

## 4. Results and Discussion

### 4.1 Nickel coated Anodized aluminum oxide membranes

Preparing a nickel-coated anodized aluminum (AAO) membrane at cryogenic temperatures for analysis in a LEAP 5000 XS is an interesting case study. AAO membranes are used for the preparation of nanostructured materials, and because of their hydrophilicity, are ideal templates to hold liquids and freeze them for subsequent APT analysis. Coating the AAO membrane with nickel can improve its hydrophilic nature, which can make it easier to load sample onto the membrane. Alumina is a ceramic, which hence presents its own challenges for FIB-based preparations, i.e. charging, low milling rates, and APT analysis.

Fig. 3(a) presents the top and Fig. 3(b) the cross-sectional view of the AAO templates showing deposited Ni to a depth of approximately 3 μm. This sample was attached to a commercial Cu clip along with a Si coupon in a dual cryo-puck and was transferred to the FIB.

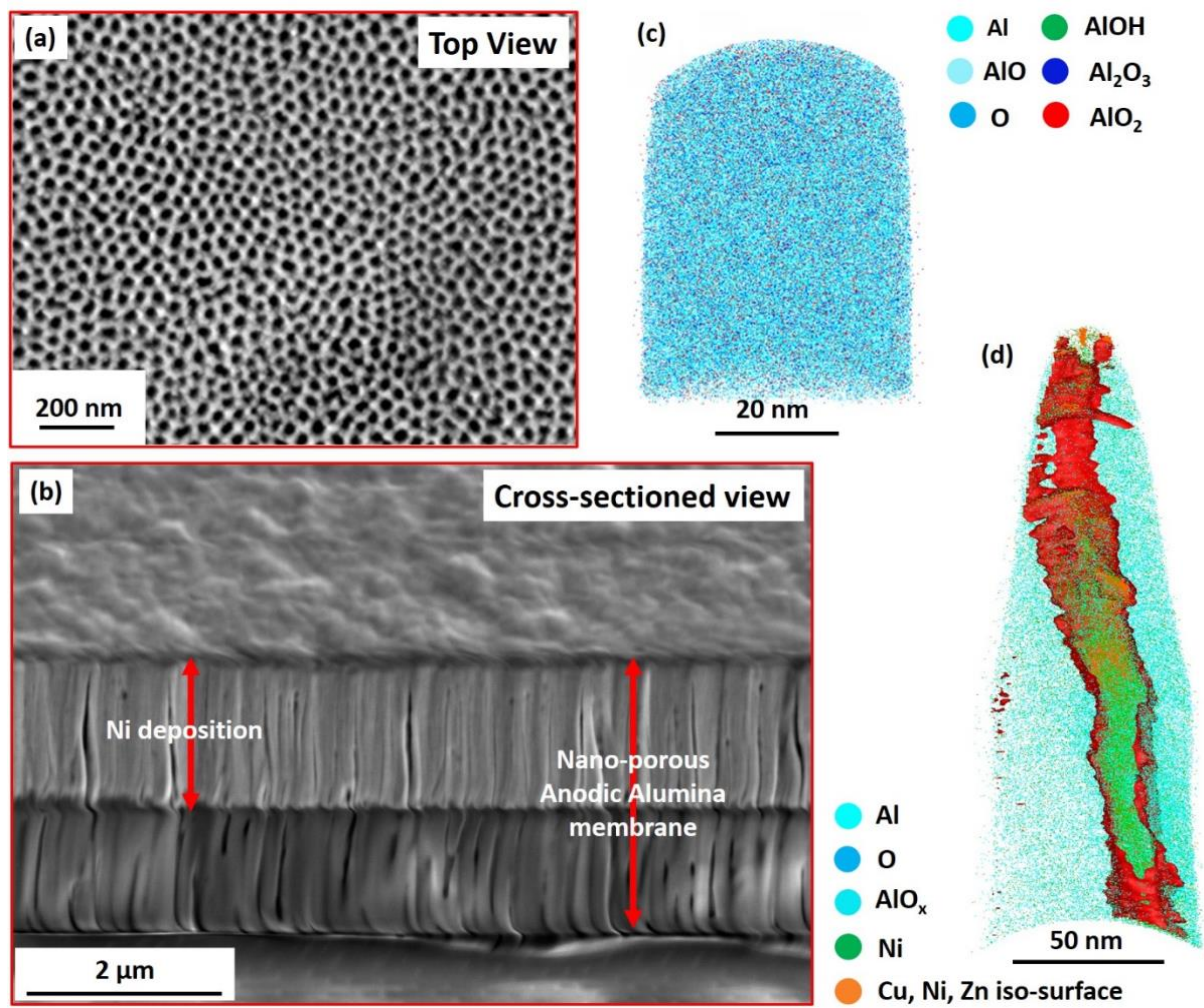

**Figure 3.** (a) Top view and (b) cross-sectioned view of typical Ni-filled pore (bottom image) (b) X-Z cross section slice of APT reconstruction of pure AAO (c) X-Z cross section of metal filled pore in AAO in APT reconstruction, visualized as 5% atomic percent metal iso-surface (Ni+Cu+Zn)

Samples were prepared as per our above-mentioned protocol at cryogenic temperatures and the transfer to the atom probe was performed under cryogenic temperatures using the suitcase. For this particular set of experiments, redeposition welding was performed with yellow brass, rather than Cr. The specimens were run in laser pulsing mode at a base temperature of 50 K, laser energy of 100 pJ, at pulse repetition rate of 125kHz with a detection rate of 1 ion per 1000 pulses on average.

A X-Z cross-section from the APT reconstruction of an analysis from the bulk AAO material, showing a homogeneous distribution of aluminum and oxygen displayed in Fig. 3(c). A filled pore in the AAO material is visualized using an iso-concentration surface of 5 at. % Ni, Cu, and Zn in Fig. 3(d). The Cu in the nickel-filled pore can be explained by the presence of small crack in that region of the AAO template, and when the APT specimen was fabricated, some brass was redeposited and integrated into the pore itself.

### 4.2 Cryo-FIB Preparation of De-zincified Brass

De-zincified yellow brass (initially Cu:63%, Zn:37%) was used as an example of metallic materials. De-zincification creates nanoporous structure which can serve as an affordable template to hold liquids for APT studies. Fig. 4(a) shows the top and cross-sectioned view of the brass after dezincification.

The Si posts and the de-zincified brass piece were mounted on copper clips in the dual puck and transferred to the FIB at room temperature. The FIB cryogenic stage was cooled to cryogenic temperatures and atom probe specimens were prepared. After preparation, the stage was brought back to room temperature. The transfer to the atom probe was done in ambient atmosphere. The measurement was performed at 50 K, with a laser pulse energy of 40 pJ at a pulse repetition rate of 200 kHz with a detection rate of 5 ions per 1000 pulses on average. Fig. 4(b) shows the (002) Cu pole in the detector event map. Slices through the APT reconstruction from the X-Y plane and X-Z planes are provided in Figs. 4(c) and 4(d), respectively. The bulk composition obtained from the whole tip suggest a Cu and Zn content of 87 and 13 at. %, respectively, indicative of the loss of Zn expected from the incomplete dezincification.

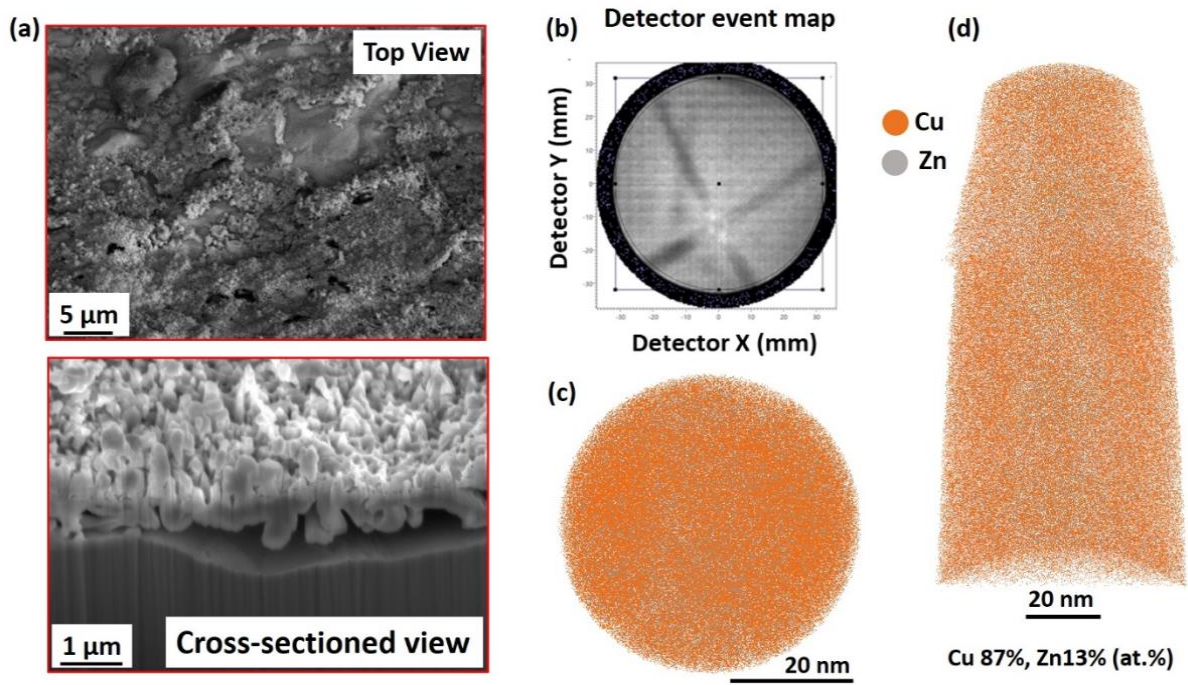

**Figure 4.** (a) Top and cross-sectioned view of Brass sample after dezincification, (b) Cu (002) pole in detector event map and APT reconstruction of low-Zinc brass, after cryogenic sample preparation (a) Y-Z slice (b) X-Y slice.

### 4.3 Frozen liquids on Cu-based substrate

Previous reports of APT analysis of frozen aqueous solutions showed the importance of selecting a hydrophilic substrate to hold the liquid of interest (El-Zoka et al., 2020; Schwarz et al., 2020). The de-zincified brass discussed above have nano-porosities which can hold aqueous solutions for freezing and APT specimen preparation. Here, a 0.1 M arginine hydrochloride solution was prepared and a droplet of 2 μl was placed on the surface using a pipette inside the glovebox and blotted as required. The dual puck, having Si micro posts on one the clip and liquid on the second clip was plunged in LN2, and subsequently transported from the glovebox into the FIB using the pre-cooled Ferrovac suitcase, using the procedure in (Stephenson et al., 2018).

Fig. 5 (a) - (f) details the corresponding preparation process. The frozen liquid on dezincified brass was lifted-out and the slice is mounted on the pre-sliced Si posts as shown in Fig. 5a. The lifted frozen liquid was mounted using redeposition cuts and the attachment with the Si post was done using redeposition cuts using current of 40 pA at 30 kV with cut dept of 1 μm.

Redeposition cuts are visible in Fig. 5a. Figs. 5(b) — (c), are top and side views of the mounted sample on Si. Fig. 5(d) shows the side cut at the interface and the Cr deposition in the cavity to strengthen the bond, while Fig. 5(e) displays the milled sample with water on top, then brass supported by the Si post. Fig. 5(f) is the final prepared APT specimen of the frozen solution at higher magnification. There is good contrast between the liquid layer and the metallic support material.

Specimens were transferred to LEAP 5000XS using the pre-cooled suitcase. Specimens were analyzed at a base temperature of 50 °K, 40 pJ laser pulse energy, at a repetition rate of 125kHz, and a detection rate 7 ions per 1000 pulses on average. The corresponding APT mass spectrum is shown in Fig. 6, and in the inset the X-Y cross-section slice through the reconstruction is presented.

The carrier substrate Cu and Zn, and their detection in the liquid, far from the substrate is likely due to their dissolution into the liquid prior to freezing. These metallic species are detected across the entire dataset and not only on the edges, as could be expected if they originated by redeposition during FIB milling. Their presence leads to peak overlap of Cu and

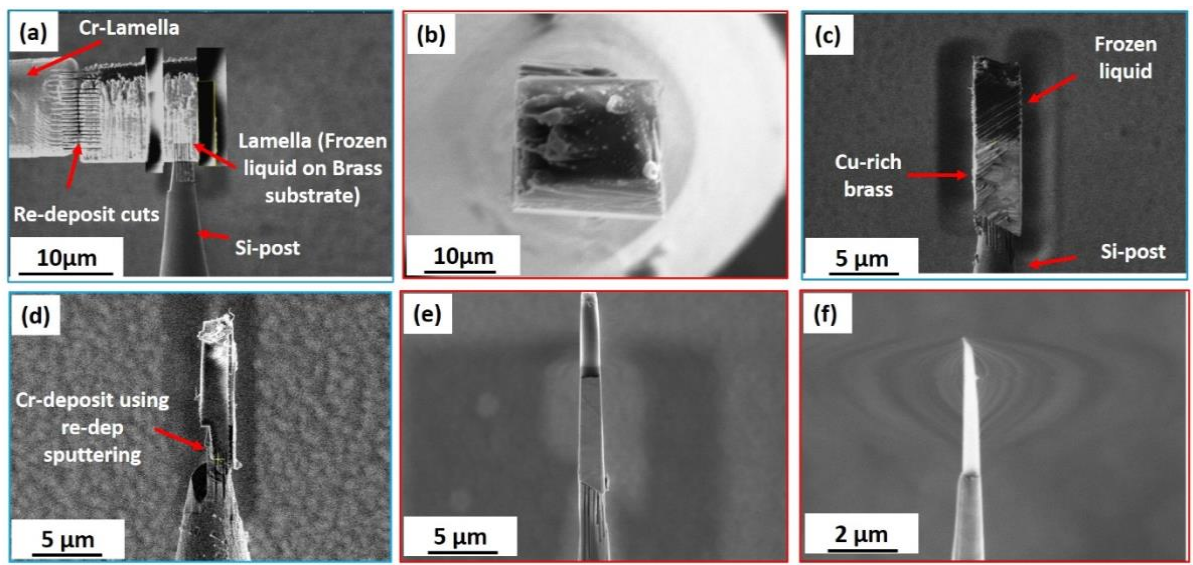

**Figure 5.** Cryogenic APT specimen preparation procedure for a 0.1M arginine HCl solution in Type I ultra-pure water on nonporous Cu. (a) FIB view, mounting ice on brass on Si microtip (b) SEM view of mounted ice sample (c) FIB cross-section view of ice APT specimen on brass (d) ion-beam image showing the cavity milled at the interface of Si post and sample, Cr is deposited in the cavity, (e) Sharpened tip at the  (f) SEM view of final sharpened ice APT specimen.

Zn with some of the organic molecules. A different substrate should be used to avoid mass interface of this particular molecule.

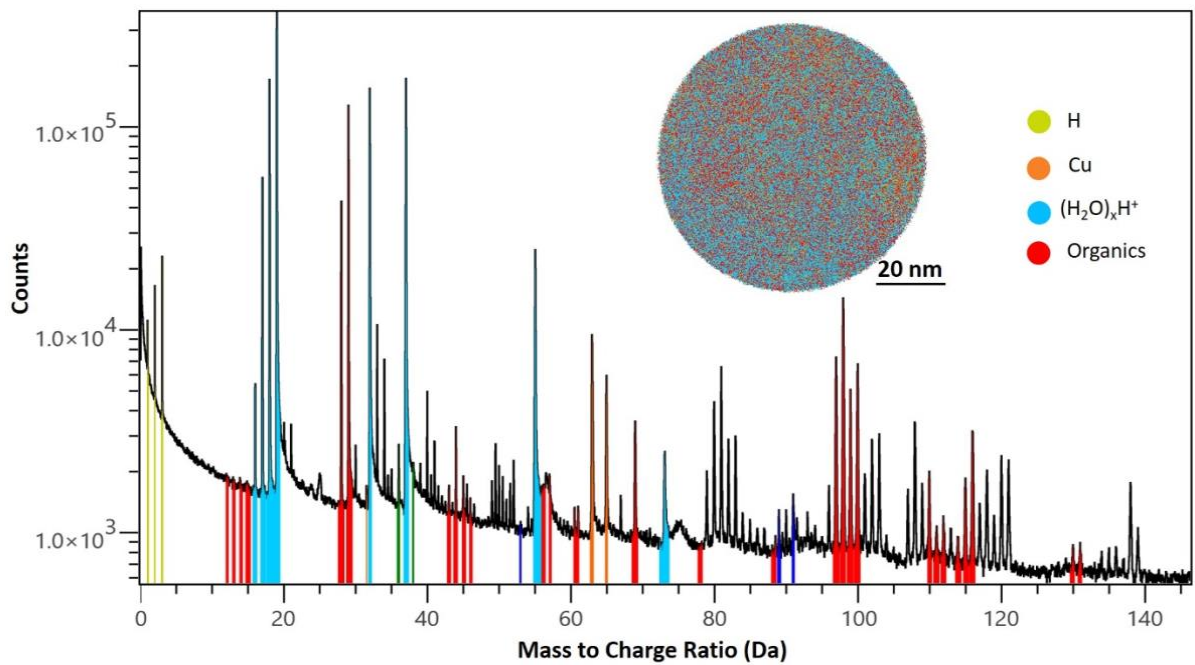

**Figure 6**. Mass spectrum from the AA solution on support (inset) selected APT ion range X-Y view.

**Direct Lift-out of bulk frozen liquid without support**

Direct, site specific lift-out of liquids for APT analysis has not been demonstrated. There are reports cryo-TEM sample preparation of frozen liquids (Mahamid et al., 2016; Schaffer et al., 2019), however those methods rely on Pt-GIS for sample coating and welding. In an attempt to lift-out only frozen liquid (i.e. without substrate) we have prepared a solution using Type I DI water and a dilute phosphate buffered saline aqueous solution. A deposited droplet of the liquid on a nanoporous Cu support was plunged into LN2, and inserted into the FIB through the suitcase. A rectangular lamella containing only the ice was cut, as shown in Fig. 7(a), then the lamella extracted and welded to a Si post using re-deposit cuts and Cr-welds as described in section 3. The close-up in Figs. 7 (b)-(c) demonstrates the absence of the substrate, with the darker areas, particularly near the Si microtip, showing lower electron contrast than the Si microtip, thereby demonstrating that the specimen is lighter and not the metallic substrate.

While annular milling is typically performed for APT specimen preparation, in this specific case it must be generally avoided. Frozen aqueous solutions mill at a much higher rate than the underlying Si post, making it challenging not to mill through the entire layer and damage the redeposition weld, which would lower yield. Here, two rectangles were set at +5° and -5° tilt, as in Fig. 7(c), and directionally milled at 80 pA then 40 pA towards the interface rather than away from it for varying time (Larson et al., 1998). The stage was then rotated 90° and the procedure was repeated, making a tall pyramidal structure. Final annular milling was performed at 30 kV, 40 pA and 24 pA, with a 5 kV 24 pA beam shower to remove any severely implantated and damaged regions, and the image of the final specimen is shown in Fig. 7(d).

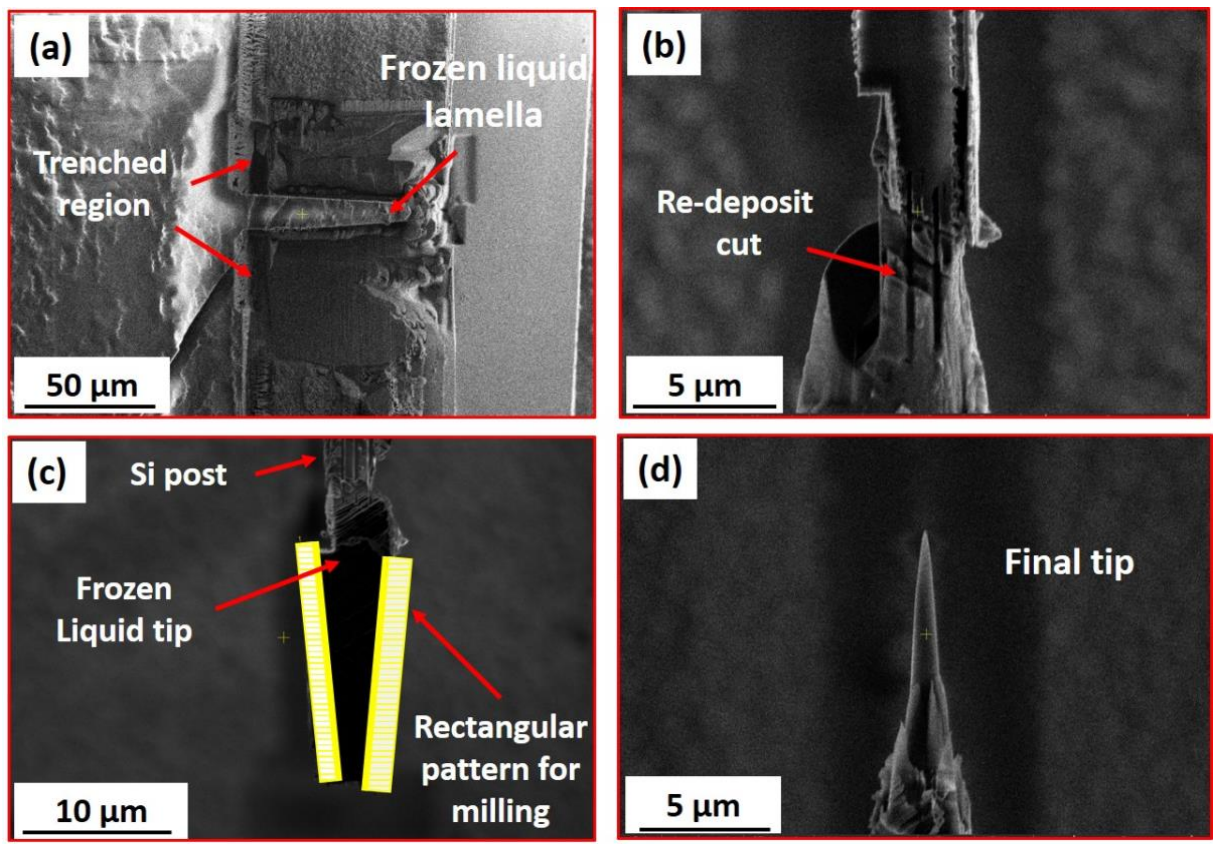

**Figure 7.** Preparation procedure for pure water lamellas. (a) Bulk ice (b) Pure water APT specimen, zoomed version of weld showing water down to Si interface (c) Wedge sharpening of APT specimen (d) final sharpened APT specimen.

Specimens were analysed in both voltage and laser mode. In Fig. 8 a mass spectrum is plotted for a dataset obtained containing approximately one million ions obtained in at a base temperature of 50 °K, with a pulse fraction of 20%, and a repetition rate of 200 kHz and detection rate of 1.5 ions per 1000 pulses on average. It contains protonated water clusters $(H_2O)_nH^+$ where n is the number of the water molecules. An example of mass spectrum, obtained in laser pulsing mode is shown in Fig. 9. It contains 7 million ions, extracted from a larger dataset, obtained at 50 K, 66 pJ laser pulse energy, at a repetition rate of 100 kHz, and a detection rate of 5.5 ions per 1000 pulses on average. The spectrum is dominated by similar albeit water clusters and a higher level of background. The higher thermal tails from the water clusters make it impossible to observe anything else in the mass spectra. In both cases, the molecular ions of water match previous work on metallic supports (El-Zoka et al., 2020; Schwarz et al., 2020), and from other groups looking at water in atom probe tomography (Panitz, 1991) or desorption and ionization studies of water from sharp specimens, e.g. field emitters, under high field conditions (Stintz & Panitz, 1991; Pinkerton et al., 1999; Stuve, 2012). Larger laser mode datasets up to 40 million ions have been collected, demonstrating that the welding method works well. The spectrum reported in Fig. 9 was from a significantly shorter APT specimen than the one shown in Fig. 7(d). It was approx. 4 µm tall (when measured from the highest edge of the Si micro post). In general, taller ice APT specimens have higher temperature rise and slower cooling, and the mass resolution and thermal tails for taller APT specimens are significantly worse. A significantly taller APT specimen, in excess of 10 µm, only showed water clusters to n=5, indicative of a larger field drop as discussed by (Stephenson et al., 2018).

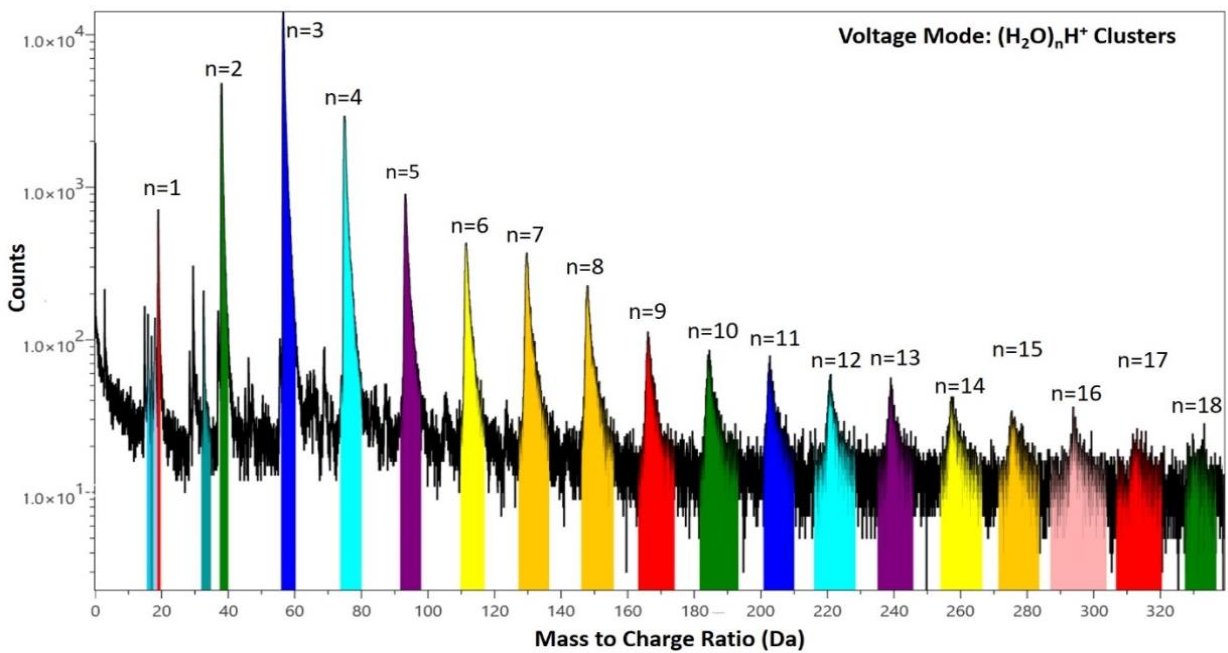

**Figure 8.** Mass-spectrum from the Voltage mode run of frozen liquid APT specimen.

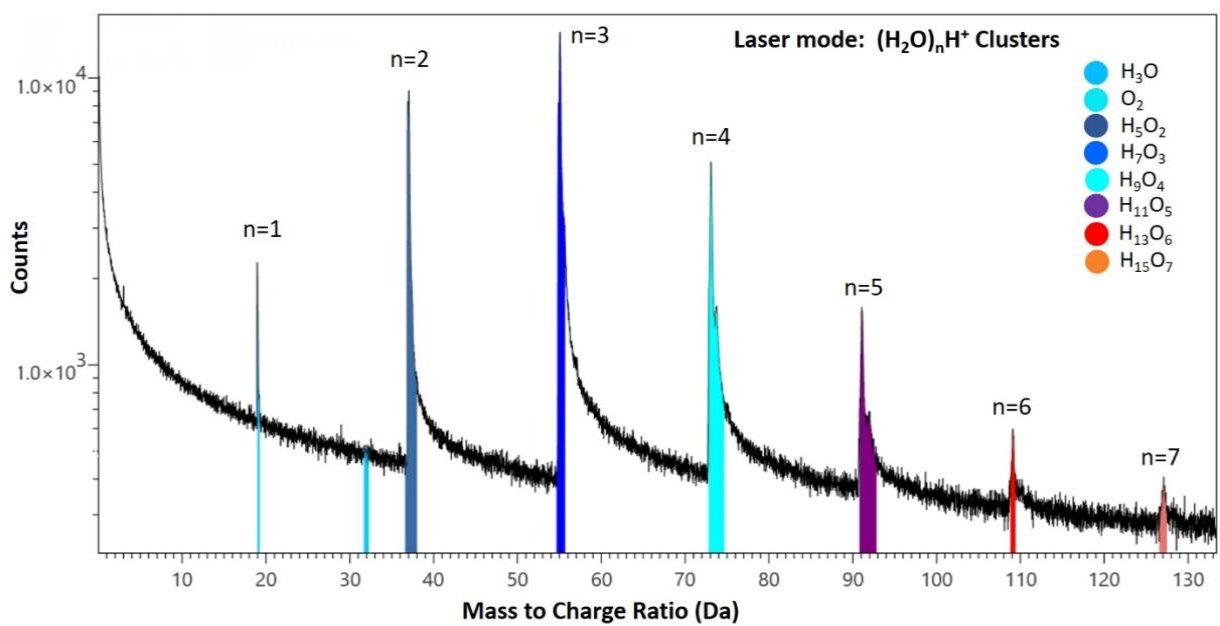

**Figure 9.** Mass-spectrum from the voltage mode run of frozen liquid APT specimen.

### 4.4 NMC811 battery cathode material

Battery materials tend to be air and beam sensitive, making specimen preparation and transport challenging. In addition, the diffusion of lithium ions driven by the electrostatic field can result inhomogeneous distribution of Li-ions (Pfeiffer et al., 2017; Greiwe et al., 2014). Kim et al. (Kim, Antonov, et al., 2022) proposed that the formation of a conductive thin layer would shield the electrostatic field conditions, that facilitates early specimen fracture.

Specimens from a commercial battery cathode material, NMC811, were prepared at cryogenic temperature. In order to enable shielding, we aimed to deposit a thin metal film on the finally-shaped specimens using method introduced by (Kölling & Vandervorst, 2009). Different geometries of the Cr-lamella for re-deposition sputtering on sharpened APT specimens, as shown in Fig. 10 a – c was tried. The results obtained from different geometries is shared in supplementary files. The most even coverage was obtained for the semi-circular geometry, as shown in Fig. 10 c and d.

The micromanipulator with the attached Cr-lamella was brought to the center of the ion-beam focus and a semi-circular (radius of 8 μm) cut was made using ion beam current of 0.23 nA at 30 kV, shown in Fig. 10(c). In the next step, the Cr-lamella with the semi-circular cavity is maneuvered such that it aligns very close to the sharpened APT specimen. Subsequently, a semi-circular pattern with inner diameter of 8 µm and outer diameter of 11 µm is placed on the cavity, and to avoid the cutting of the prepared specimen, a rectangle with disabled milling is placed alongside the circular pattern. The milling was performed inside out with a current of 40 pA for 20 s. After that the Cr-lamella was retracted and stage was rotated by 90º degree, and the same procedure was repeated to cover the other sides of the APT specimen. To ensure proper coating, sputtering was performed from all 4 sides. The stage along with coated APT tips were brought to room temperature.

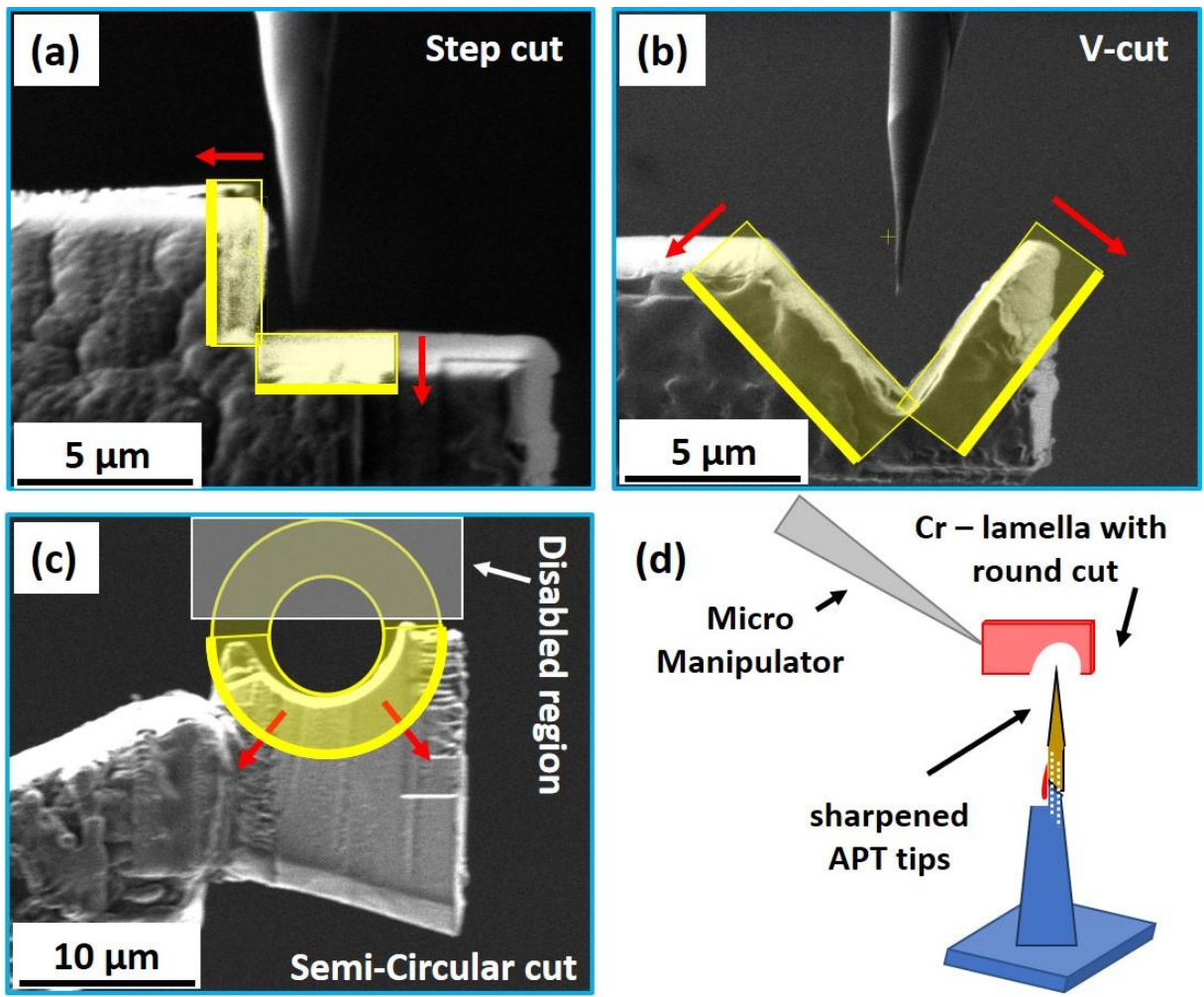

**Figure 10**. Ion-beam Image for different geometries tried for re-deposit geometries (a) Step geometry, (b) V-cut geometry and (c) semi-circular cut geometry, (d) a schematic for the arrangement of the tip across the circular cut. Pattern used for re-deposit sputtering is illustrated in yellow and milling direction is marked with red arrows.

Fig. 11 shows a secondary electron micrograph of secondary particle from the NMC811 powder. In-Situ Cr-coated APT specimens were prepared from this particle, using the abovementioned methodology and transferred into the atom probe through the suitcase. Fig. 11b shows the mass spectrum from an APT analysis at a base temperature of 60 °K, a laser pulse energy of 30 pJ, a pulse repetition rate of 125 kHz, and a detection rate of 8 ions per 1000 pulses on average, shows clear peaks associated to Cr. APT reconstruction of the sample is presented in Figs. 11(c) and (d), shows a thin layer of Cr surrounds the imaged region. No signs of heterogeneous Li distribution associated to in-situ delithiation are observed in this dataset, despite the relatively high laser pulse energy used to compared to reports by (Kim, Antonov, et al., 2022). Note that the coating technique covers some but not all of the specimen's length, particularly if these are several µm in length. Once the atom probe analysis moves into regions in which specimen is uncoated, thermally induced lithium migration was seen to occur.

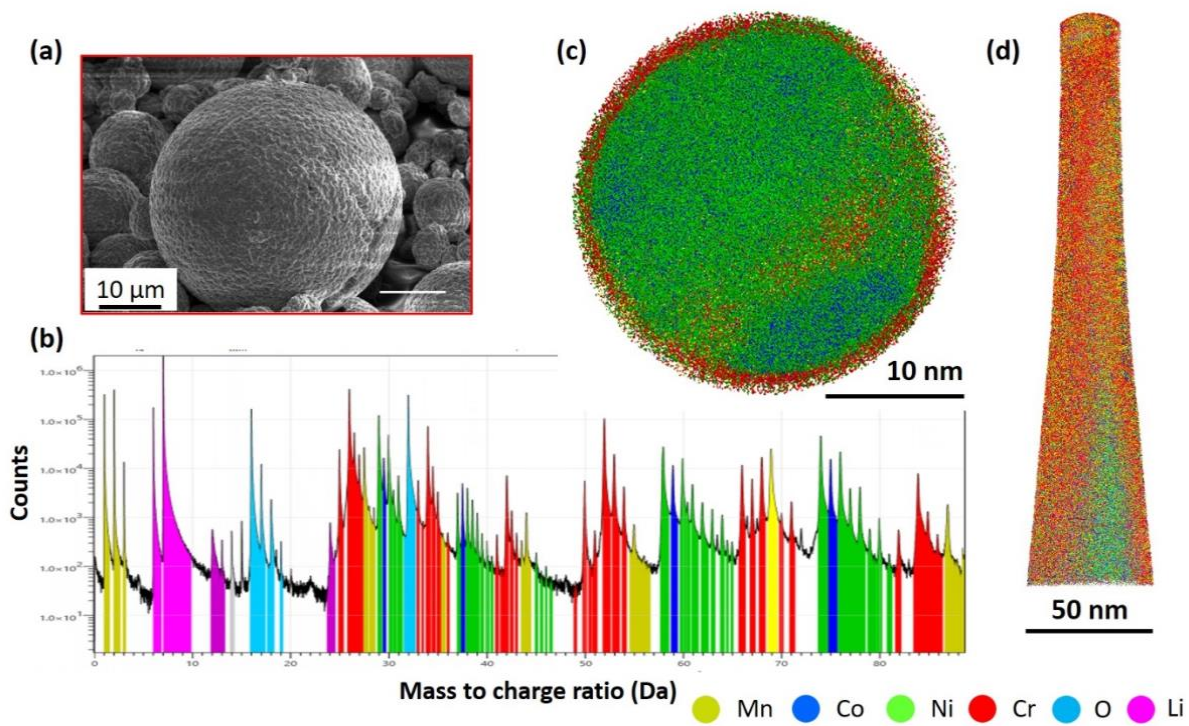

**Figure 11.** (a) Secondary electron image of the NMC811 cathode powder sample, (b) mass-spectrum obtained from APT run of the NMC tips, (c) APT reconstruction of the 4-sided coated NMC tips and prepared using circular cuts on Cr-lamella (image is from the X—Y slice), (d) visualization of the same tip from Y—Z plane.

## Conclusions and Future work

Eyeing the future goals of enabling APT analyses for frozen wet samples, we demonstrate an efficient, reliable and likely universal cryo-FIB-lift-out sample preparation technique for APT studies. Expanding the use of FIB-based specimen preparation at cryogenic temperatures will also enhance the quality of APT data. As an added advantage, the current workflow allows multiple APT specimens from a single lift-out from a selected ROI.

We showcased application of this protocol to a wide range of samples, including metal, oxides, liquids and beam sensitive battery materials, for preparation of specimens suitable for APT analysis.

Attaching specimens by redeposition cuts proves to be sturdy and efficient, potentially overcoming the issues with Pt- based-GIS at cryogenic temperatures. Re-deposition cuts can be performed both at room and cryo temperatures.

We demonstrated cryo-lift-out of frozen liquids with and without substrate using re-deposition procedure. The prepared specimen was also analyzed by APT in both voltage and laser pulsing mode. The quality of the obtained mass spectra was encouraging, but further optimization is required in future.

Specimen from the battery cathode materials were successfully prepared, coated with Cr from all sides at cryogenic temperatures. The corresponding APT results suggest that beam damage was avoided by cryo-preparation and delithiation was avoided using Cr-coating.

In the future, we aim to perform site-specific grain boundary preparation for instance. Optimization of coating material and thickness could be performed. With respect to direct lift-out of liquids, future work includes optimizing APT specimen length and shape to reduce thermal tail, as well as the sue of different metals for the weld and using sputter coating to improve heat and electrical conductivity, which currently remain a significant problem.

## Conflicts of Interest

There are no conflicts to declare.

## Acknowledgments

EVW, SHK, AEZ, and BG are grateful for funding from the ERC for the project SHINE (ERC-CoG) #771602. FG and BG are grateful for funding from the EPSRC under the grant #EP/V007661/1. TMS and BG is grateful to the DFG for funding through the Leibniz Award. We thank Uwe Tezins, Christian Broß and Andreas Sturm for their support at the FIB and APT facilities at MPIE. We would like to thank Katja Angehangt and Monika Nellessen for her assistance with sample preparation.